\documentclass[prb,floatfix,showpacs,12pt]{revtex4}
\usepackage{graphicx}

\begin{document}

\title{ Phase Effects on the Conductance Through Parallel Double Dots}
\author{ V.M. Apel$^1$, Maria A. Davidovich$^1$, G. Chiappe$^2$ and E.V. Anda$^1$}
\affiliation{$^1$Departamento de F\'{\i}sica, Pontif\'{\i}cia
Universidade Cat\'olica do Rio de Janeiro, 22453-900,  Brazil \\
$^2$Departamento de F\'{\i}sica,  Facultad de Ciencias Exactas
y Naturales, Universidad de Buenos Aires, 1428, Argentina}

\begin{abstract}
Phase effects on the conductance of a double-dot system in a
ring structure threaded by a magnetic flux are studied. The Aharonov-Bohm
effect combined with the dot many-body charging effects determine the
phases of the currents going through each arm of the ring. The
cases for zero magnetic flux or half a quantum of flux are discussed
in detail. It is shown that, depending upon the magnetic flux and
the state of charge of the dots, controlled by gate potentials,
the dephasing of the upper and lower arm current gives rise to a
$S=1/2$ or $S=1$ Kondo regime. We also show that even in the
absence of a magnetic flux there can be a circulating current in
the ring, depending on the system parameters.
\end{abstract}
\pacs{73.63.-b,73.63.Kv}
 \maketitle

In the last years attention has been focused on the properties of
double-quantum-dot systems\cite{Various} since they are expected
to be basic building blocks for quantum computing\cite{DiVicenzo}.
In this context mesoscopic coherent transport is a key
 phenomenon that can probe entanglement\cite{Loss} and phases accumulated by electrons 
traversing the system.
 The phase-coherent transport effects can be analyzed by embedding the dots into an 
Aharonov-Bohm ring threaded by a
 magnetic flux and connected to leads.  Theoretical\cite{variousT} and experimental
\cite{variousE} works have discussed the transmission phase shifts of single and double  
quantum dot systems in the Coulomb blockade regime.
More recently the effect of Kondo correlation on the transmission phase of a quantum dot has 
been measured\cite{Heiblum}
 and theoretically discussed\cite{Aharony,Hof,ANDA,KANG,ECKLE,YOUNG}.

The two-quantum-dot system is particularly interesting since, when each dot is inserted into 
one arm of a ring connected
to leads, as represented in Fig. 1, it presents two paths for the electrons to go through, 
producing interferences that
depend upon the phase in each arm. The effect of the interferences on the spectral densities 
of such a system has been
studied\cite{Boese}, taking the intra and inter-dot Coulomb repulsion to be infinite. This 
limit restricts the study of
the Kondo phenomenon and the phase shift interference effects to a situation in which the 
number of electrons in the system cannot be greater than one.
However, depending on the magnetic flux enclosed by the ring, new and very interesting physics 
appear when both dots are charged and
in the Kondo regime. Moreover, the general problem of the transmission phases in this system 
as the magnetic flux and
the state of charge of the dots are varied has not been yet clarified. The dephasing between 
the two contributions to
the current do not depend exclusively upon the state of charge of each dot, as a naive 
interpretation of the Friedel
sum rule could predict. This is a consequence of the fact that the phases associated to the 
arms are not independent
objects because one is renormalized by the other\cite{Aharony,Hof}.

The purpose of this letter is to contribute to the understanding of this problem.
We discuss phase effects on the conductance of this system using an exact numerical 
diagonalization algorithm that provides only the ground state of the system. Our study is therefore restricted to zero 
temperature.
This procedure permits us to take into account in an adequate way the interference effects due 
to the interaction between the two ring arms.
Different Kondo regimes are accessed by varying the magnetic flux and the charge in the dots 
controlled by the gate potentials. The Aharonov-Bohm effect
combined with the dot many-body charging effects determine the phases of the currents going 
through each arm of the ring.
In order to have confiability in the numerical results, we adopt a dot-lead interaction
such that the Kondo cloud is of the order of the cluster we exactly diagonalize\cite{Guil}. In this case 
the plateau shape of the conductance, when the system is in the Kondo regime, is more like a Lorentzian.  
The conductance shows quite different behaviors for the magnetic flux $\Phi =0$ and $\Phi=\Phi_0/2$. In the first case
the dots act coherently and the system is in a Kondo state of total spin $S\sim 1$ while in 
the second, the dots are
uncorrelated and in a $S={1\over 2}$ Kondo regime. For other fluxes the conductance shows 
intermediate behavior
between these two extremes. As an important consequence of interference we also find that 
current can circulate around the ring even in the absence of magnetic fluxes.

An Anderson two-impurity first-neighbor tight-binding Hamiltonian
represents the system shown in Fig. 1,
\begin{eqnarray}
H &=& \sum_{ r=\alpha,\beta\atop\sigma}\left( V_{r} +\frac{U}{2} n_{r \bar\sigma}\right)
n_{r\sigma}+  t \sum_{i,j}c^+_{i\sigma}c_{j \sigma} \label{H}\nonumber\\
&+& t'\left[e^{i{\pi\over 2}{\Phi\over \Phi_0}} (c^+_{\alpha\sigma} c_{1\sigma}\! +\!c^+_{1 \sigma} c_{\beta \sigma}\!+\! c^+_{\beta\sigma} c_{\bar1\sigma}\! +\!c^+_{\bar1 \sigma} c_{\alpha \sigma}\!)+\!{\rm c.c.}\right]          \nonumber\\
 \end{eqnarray}
where $V_\alpha$ and $V_\beta$ are the gate potentials applied to
the dots, $U$ is the Coulomb repulsion, considered to be equal for
the two dots and $\Phi$ is the magnetic flux threading the ring.
The parameters t' and t are, respectively, the hopping matrix
element between the dots and their neighbors and between sites in the leads.
The one particle Green functions $ G $ are imposed to satisfy a Dyson equation $\hat G = \hat g + \hat g \hat T \hat G$ where $\hat g$ is the Green function matrix of a
cluster containing the ring with the dots and a number of atoms of
each lead and $\hat T$ is the matrix Hamiltonian that couples the
cluster to the rest of the system. The undressed Green function $
\hat g $ is calculated using the cluster ground state obtained by
the Lanczos method. This approximation\cite{Guil} has shown to be very
accurate when the cluster is of the size of the Kondo cloud
$hv_F/T_K$, where $v_F$ is the Fermi velocity and $T_K$ the Kondo temperature, although it gives
qualitatively reliable results even for smaller clusters, compatible with
the Friedel sum-rule and the Fermi liquid properties of the system.

The transport properties are studied as a function of the gate potentials 
applied to the dots. Taking the energies in units of the Coulomb
interaction $U$, we set the leads bandwidth $W=8$ and $\Gamma=t'{^2}/W=0.05$.
The Fermi level is chosen to be $\epsilon_F = 0$.
The conductance $G=dI/dV$ is obtained using the Keldysh formalism to calculate the total current 
$I=\sum_{ i=\alpha,\beta}I_i$, where  
$I_{\alpha}$ and $I_{\beta}$ are the contributions coming from the two ring arms given by,
\begin{eqnarray}
I_{i}&=&\frac{2et'^{2}}{h}\int_{-\infty }^{\infty } \left[(g^{-+}_{L}(\omega)-g^{-+}_{R}(\omega))\times Im\left\{G^r_{\bar 1\bar
1}(\omega)G^a_{\bar1 i}(\omega) - G^r_{\bar1 1}(\omega) G^a_{1i}(\omega)
\right\}\right]~d\omega 
\end{eqnarray}
where sites $1$ and $\bar 1$ are the dots first
neighbors, $G^{r}$ and  $G^{a}$ are the dressed retarded and
advanced Green functions and $g^{-+}_{L}$ and $g^{-+}_{R}$ are,
respectively, the density of states at the left and right contacts
multiplied by the Fermi distribution function. The Green functions are calculated at the Fermi level.

The conductance for two values of the magnetic flux, $\Phi=0$ and $\Phi_0/2$, is represented 
in Fig. 2 as a function of the energy of the dot levels. It results to be
weakly dependent upon the magnetic flux in the regions of the parameter space where only one 
dot is active so that
current flows essentially along one arm of the ring. It possesses the characteristics of the 
one-dot
conductance with a width, as a function of the gate potential, of the order of $U$ due to the 
Kondo effect of the
charging dot. On the other hand, when $V_{\alpha}\sim V_{\beta}$, the conductance is strongly 
dependent upon the
magnetic flux, showing special different features for the cases $\Phi=0$ and $\Phi_0/2$. Let 
us focus on the region
$ -1 < V_{\alpha},V_{\beta} < 0$ of Fig. 2. The differences in the conductance between these 
two cases are more
striking when the dot level energies are similar, $\Delta E \sim 0$ (diagonal continuous line). In this case the two arms are at resonance since both dots are in the
Kondo regime. For $\Phi=0$ the two arm currents are in phase and interfere constructively and the 
conductance has one broad peak. As the 
magnetic flux is turned
on, the currents along the two arms are no longer in phase and the transport properties change 
qualitatively.
For $\Phi = \Phi_0/2$, the current arm amplitudes are out of phase and the conductance for 
$\Delta E=0$ cancels out for
all values of the gate potentials, as shown in Fig. 2b.

For small values of $\Delta E$, the conductance presents two peaks for $\Phi =\Phi_0/2$, and 
three for $\Phi=0$, the
most interesting case. The spin-spin correlation, the charge at the dots and the conductance 
are displayed in Fig. 3 as
a function of $V_{\alpha}$ for $\Delta E=0.6 (V_{\alpha} = V_{\beta} - 0.6)$, corresponding to 
the dashed
line in Fig. 2. We first analyze the case $\Phi = 0$. As  $V_{\alpha}$ decreases from the value
 $0.5$ and charge begins
to enter into dot ${\alpha}$ its spin gets negatively correlated to the conduction electrons. 
The dot is in the Kondo
regime and the conductance increases up to a quantum of conductance maximum around 
$V_{\alpha}=-0.2$.
At $V_{\alpha}=-0.5$ the conductance cancels out and the various spin-spin correlations change 
abruptly. In the region
$-0.5 < V_{\alpha} < -1.1$ the Kondo correlation of both dots, 
$\langle\vec S_{\alpha} \vec S_{c}\rangle$ and $\langle\vec
 S_{\beta} \vec S_{c}\rangle$,
gets stronger (Fig. 3b) and the conductance grows again reaching a second maximum at the
electron-hole symmetry condition, $V_{\alpha}=-0.8$, where the system has just two electrons.

The most interesting aspect of the conductance is that its lateral peaks are due to a Kondo 
regime that is different
from the one of the central peak. This conclusion can be obtained from the analysis of the 
spin-spin correlation of the
two dots, shown in Fig. 3b. When the gate potential is reduced from $V_{\alpha}=0.5$, dot 
$\alpha$ enters into resonance and
into the Kondo regime, while the other is well above the Fermi level and empty  of charge. The 
current circulates only
along dot $\alpha$ giving rise to the first peak of the conductance. It is a typical  $S=1/2$ 
Kondo effect. As the gate potential is further reduced the electronic charge begins to drop also 
into dot $\beta$, giving rise to a new Kondo peak that, due to level
repulsion, pushes the other one, at $\alpha$, to values below the Fermi
level. This reduces the DOS at the Fermi level, diminishing the conductance of the system, 
and raises the dot $\alpha$ charge to a value above unity (Fig. 3c), increasing the energy of 
the ground state due to Coulomb interaction. As a consequence, a level crossing takes place 
and an excited state with total dots spin mean value $<S_T> \sim 1$ and less charge at dot 
$\alpha$ becomes the new ground 
state of the system. This process gives rise to an abrupt reduction of the dot charge and to 
a restoration of the Kondo resonance at the Fermi level.
While in the gate potential region corresponding 
to the original ground state the two dots are weakly antiferromagnetic correlated, in this 
new 
ground state they adopt a coherent behavior and have a strong ferromagnetic spin-spin 
correlation, $\langle\vec S_{\alpha} \vec S_{\beta}\rangle$, independent of the gate 
potential, as shown in Fig. 3b. The system is in a $S=1$ Kondo regime.
The two crossing states have opposite parity. The 
crossing itself and the consequent abrupt change of the 
physical properties is derived from the fact that the system is spatially invariant under 
reflection so that parity is a good quantum number. When this symmetry is broken by 
introducing a small magnetic field (note that we are analyzing the case $\phi = 0$) or 
an asymmetric perturbation to the Hamiltonian, the ground state results to be a linear 
combination of these two previous states. In this case the level crossing transforms into  
a cross-over behavior and the abrupt changes of the physical quantities are eliminated.
This phenomenon is enhanced when the interaction among the dots and 
the conduction electrons is increased. 
When this interaction is large enough, the Kondo cloud is reduced to the size of the 
cluster we exactly diagonalize. In this region of the parameter space our approximation is 
almost numerically exact and independent of the cluster size we adopt, for sizes greater than the Kondo cloud.

The physics of the $\Phi=\Phi_0/ 2$ case is diverse from the previously discussed. In this 
situation the dot spins are
weakly correlated while each dot is independently Kondo correlated to the electrons of the 
leads, as appears in
Fig. 3e.
This different behavior can be clarified by a perturbation theory argument. The effective 
correlation between the dots
can be obtained by taking the non-diagonal matrix elements that connect the dots to the rest 
of the circuit, $t'$, as a
perturbation. Due to the system topology it is clear that to get the dominant contribution to 
the effective inter-dot
interaction it is necessary to go to forth order in perturbation theory. In this case, while 
for $\Phi=0$ the
contributions that go from one dot to the other and return along the same path sum-up with the 
circulating
contributions, for $\Phi=\Phi_0/ 2$ these two contributions, having opposite signs, tend to 
cancel each other, giving
rise to a weak interdot correlation. For small $\Delta E$ the conductance(Fig. 3d) possesses 
only one peak that develops into two as $\Delta E$ increases. The charge at the dots has a 
smooth dependence upon the gate potential, as shown in Fig. 3f.  

In order to better understand these behaviors, we have calculated
the conductance of each arm of the ring, using equation (2), as well as the phases of the electronic wave functions related to these two paths.  
The conductances are displayed in Fig. 4 as a function of the gate potential $V_{\alpha}$. For 
$\Delta E = 0$ and $\Phi = 0$
(Fig. 4a) the two arms are identical and their currents circulate along the ring in opposite 
direction summing up to
constitute the total current in the leads. However, if $\Phi=\Phi_0/ 2$ (Fig. 4b) the two arm 
currents circulate in the
same direction giving rise to a persistent current along the ring and a total zero conductance.
 Satisfying a well known
property, the circulating current changes sign as the 
total charge of the ring
varies, as shown in the figure. The same phenomenon shows up 
in the case
$\Phi={\Phi_0}/ 2$ and $\Delta E = 0.6$, presented in Fig. 4d.

However, the existence of a magnetic flux crossing the ring is not a necessary condition to 
obtain a circulating current,
as can be seen in Fig. 4c, for $\Phi = 0$ and $\Delta E = 0.6$. When the two dots are unequally charged there is a
region of the gate potential parameter for which a net current circulates around the ring. 
This is due to an
interference effect among the charges going along the two arms. For other values
of the gate potentials, both currents are positive and the circulation disappears.

The quantum interference among the electrons are studied by calculating the phase and amplitude of the transmission  
that is proportional to the square modulus of $G^r_{\bar 1,1}=g_{c}t'(G^r_{\alpha,1}+ G^r_{\beta,1})$, where $g_c$ is 
the retarded Green function of the disconnected 
lead at the site $\bar 1$ and the two terms represent the contributions coming from the two arms of the system.
Taking the contribution from arm $\alpha$ as a reference, the total transmission can be 
written as $ |A_{\alpha}+A_{\beta}e^{i\phi_{\alpha,\beta}}|^2$,
where $ A_{\alpha}$ and $ A_{\beta}$ are, respectively, proportional to the modulus of $G^r_{\alpha,1}$ and 
$G^r_{\beta,1}$, and $\phi_{\alpha,\beta}$ is their phase  difference. 
It is important to notice that
$G^r_{\alpha,1}$ and $G^r_{\beta,1}$ incorporate all the renormalization of one arm due to 
the existence of the other. As a consequence, the phase of each trajectory is a function not only of its own dot charge but 
also of the charge of the
other dot, and depends as well upon the topology of the whole system \cite{Aharony}. According 
to the Onsager relation the conductance of our system, that possesses a closed geometry, is an even function 
of the applied magnetic
flux threading it, which implies $\phi_{\alpha,\beta} = 0$ or $\pi$, depending upon the gate potentials 
applied to the dots. The transmission amplitudes and the phase difference are
shown in Fig. 5 as a function of the gate 
potential for the case
$\Phi=0, \Delta E = 0.6$. When the transmission amplitudes are equal, $ A_{\alpha} = A_{\beta}$, 
the conductance cancels or has a maximum depending upon the phase difference being $\phi_{\alpha,\beta} = \pi$ ($V_{\alpha}= -0.5$ and $-1.1$)
or $\phi_{\alpha,\beta} = 0$ ($V_{\alpha} = -0.8$).

In summary we have studied a double-dot system in a ring threaded
by a magnetic flux and connected to leads. The cases $\Phi = 0$ and $\Phi_0/ 2$ are discussed. It is shown that the dephasing of the upper and lower
arm current determines whether the system is in a $S=1$ Kondo
regime, when the two dots behave coherently, or in a $S=1/2$ Kondo
regime, when they are uncorrelated. We show as well that,
depending on the parameters, a circulating current appears even if
there is no magnetic flux threading the ring. This very
interesting phenomenon will be discussed in detail somewhere else.

We acknowledge FAPERJ, CNPq, CIAM (CNPq), CAPES(CAPG-BA,12/02) and CONICET for financial support. G.C. also
acknowledges the support of the Buenos Aires University (UBACYT x210) and Fundacion Antorchas.

\newpage

\begin{figure}
\includegraphics[width=8cm]{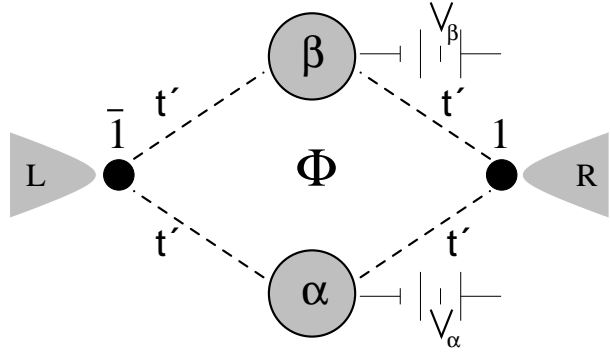}
\caption{Aharonov-Bohm interferometer with two embedded quantum dots,
 threaded by a magnetic flux $\Phi$, connected to leads.}
\end{figure}

\begin{figure}
\includegraphics[width=8cm]{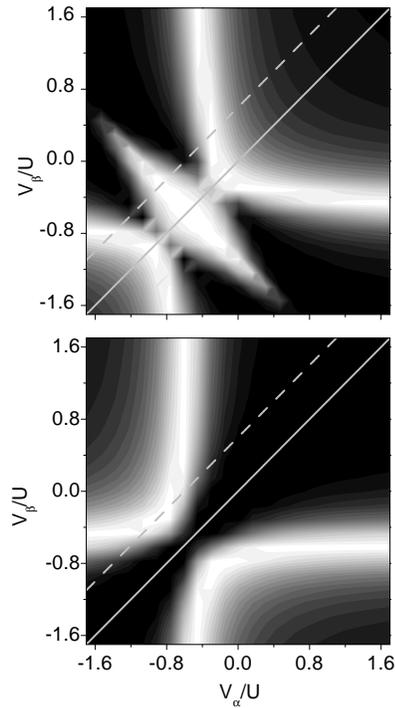}
\caption{Conductance (white, maximum; black, minimum) as a
function of the gate potentials at the dots, $V_\alpha$ and
$V_\beta$. $\Phi=0$(upper panel), $\Phi=\Phi_0/2$(lower panel);  
$\Delta E = 0$ (continuous line), $\Delta E = 0.6$(dashed line). }
\end{figure}

\begin{figure}
\includegraphics[width=8cm]{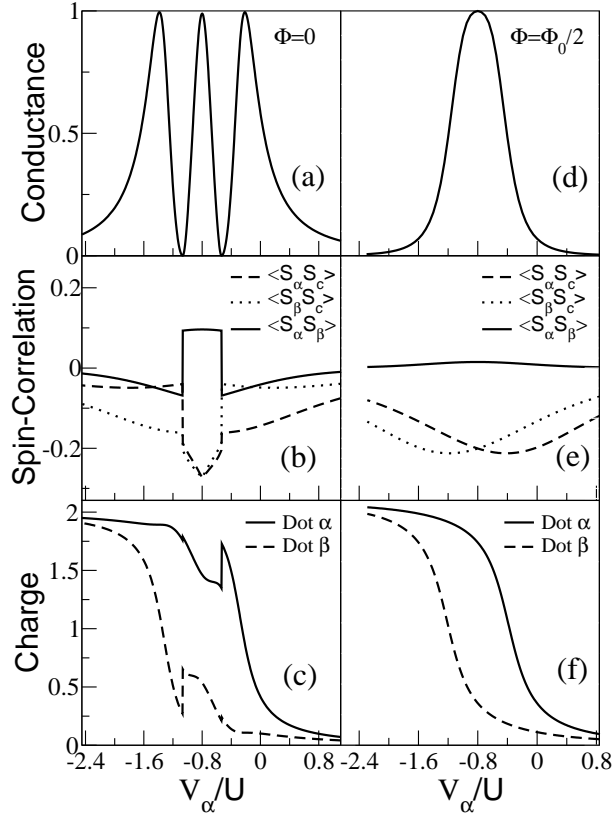}
\caption{Conductance (in units of $2e^2/h$), spin correlation and charge (in units of e) at the dots for
$\Delta E = 0.6$ ($V_{\beta} = V_{\alpha} + 0.6$) and $\Phi=0$ (left) and
$\Phi=\Phi_0/2$ (right), as a function of $V_\alpha$. }
\end{figure}

\begin{figure}
\includegraphics[width=8cm]{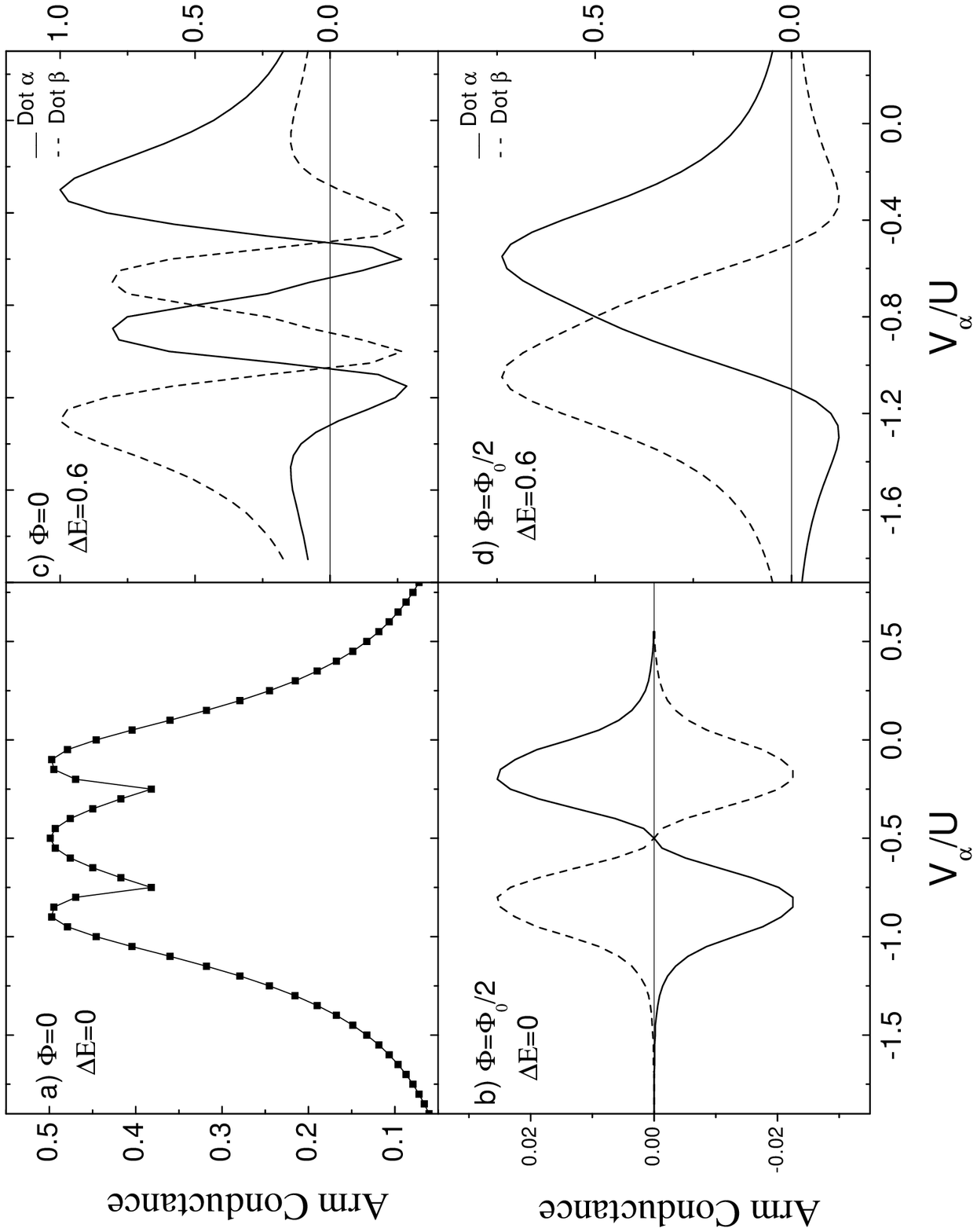}
\caption{Conductance (in units of $2e^2/h$) of the ring arm that contains dot $\alpha$ (solid
line) and dot $\beta$ (dashed line), as a function of $V_\alpha$ 
for $\Delta E = 0.$(left) and $\Delta E = 0.6$(right) and magnetic fluxes $\Phi=0$(upper) and
$\Phi=\Phi_0/2$(lower).}
\end{figure}

\begin{figure}
\includegraphics[width=8cm]{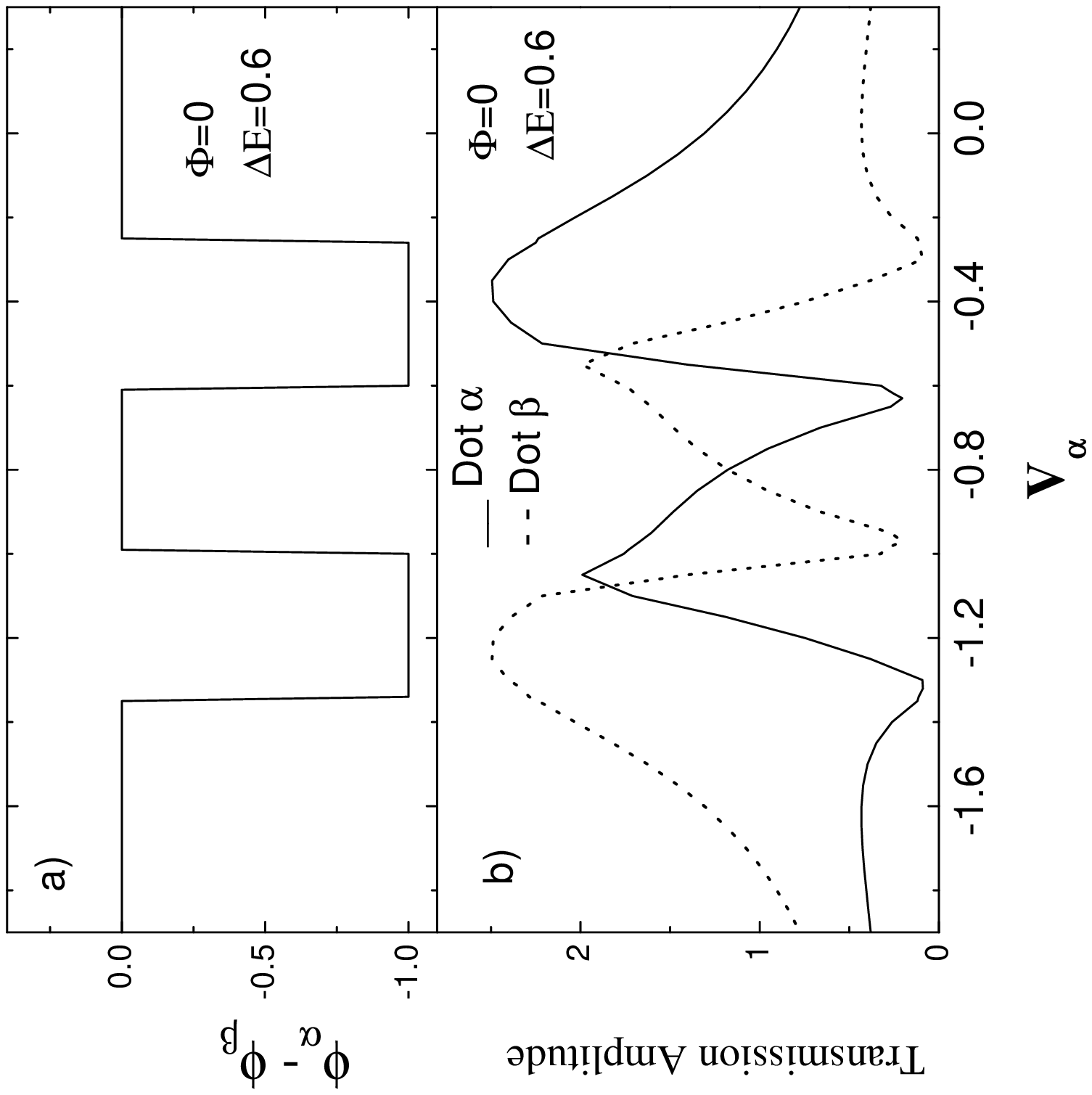}
\caption{Transmission through the two ring arms as a function of
$V_\alpha$, for $\Phi=0$ and $\Delta E=0.6$. (a) phase difference (units of $\pi$); (b) modulus(arb.units).}
\end{figure}

\end{document}